\documentclass{ragtime} 
\title[TDE and the relativistic spectral lines near SMBH]%
      {Tidal disruption events from a nuclear star cluster as possible origin of transient relativistic spectral lines near SMBH}

\author[V. Karas, M. Dov\v{c}iak, D. Kunneriath, W. Yu, 
        \& W. Zhang]  
       {Vladim\'{\i}r Karas\at{1,a} 
        Michal Dov\v{c}iak\at{1} 
        Devaky Kunneriath\at{1}      \splitauthors
        Wenfei Yu\at{2,b}    
         and                                  
        Wenda Zhang\at[]{2} \\
        \ins{1}Astronomical Institute,
        Academy of Sciences of the Czech Republic,\splitins[1]
        Bo\v{c}n\'{\i} II 1401, CZ-141\,00 Prague, Czech Republic\\
        \ins{2}Shanghai Astronomical Observatory, and \splitins[1]%
        Key Laboratory for Research in Galaxies and Cosmology,\splitins[1]%
        Chinese Academy of Sciences,
        80 Nandan Road, Shanghai, 200030, China\\
        \ins{a}\Email{vladimir.karas@cuni.cz},%
        \ins{b}\Email{wenfei@shao.ac.cn}} 

\begin{document}
\begin{abstract}
We discuss a possibility that a tidal disruption event near a dormant supermassive black hole (SMBH) can give rise to spectral features of iron in 6--7 keV X-ray signal: a relativistic line profile emerges from debris illuminated and ionised by an intense flash produced from the destroyed star. This could provide a unique way to determine parameters of the system. 

We consider a model where the nuclear stellar population acquires an oblate shape (i.e., a flattened distribution) in the inner region near a supermassive black hole, and also the primary irradiation flare is expected to occur more likely near the equatorial plane, co-planar with the infalling material. This suggests that the reprocessing of primary X-rays results in a transient profile that should be relevant for tidal-disruption events (TDE) in otherwise under-luminous (inactive) galactic nuclei, i.e. with no prior accretion disc. 

Resonance mechanisms of the stellar motion can increase the orbital eccentricity for some stars in the nuclear cluster and help to bring them close to the tidal radius, where they can give rise to TDEs. The proposed scenario appears to be distinguishably different from the standard scheme of lamp-post model for the origin of the relativistic line in active galaxies, where the source is thought to be located predominantly near the symmetry axis.
\end{abstract}

\section{Introduction}\label{intro}
Near a supermassive black hole (SMBH) tidal disruptions occur during close encounters when a plunging star on an eccentric orbit reaches the critical (tidal) radius $R=R_{\rm{t}}\sim10^{11}\left(M_\bullet/M_*\right)^{1/3}\left(R_*/R_\odot\right)$ cm, where $M_*$ and $R_*$ denote the mass and the radius of the satellite star, $M_\bullet$ is the SMBH mass \cite{evans89,luminet85,rees88}. Stars approach the event horizon, and at a certain moment they become disrupted by tidal forces of the SMBH, producing a bright flash of intense radiation that illuminates the surrounding interstellar medium and a temporary accretion disc or a ring formed by the debris \cite{cannizzo90}. According to the standard picture of galactic nuclei, we can imagine a nuclear cluster around a galactic core as a system consisting of a central SMBH, an accretion disc and a dense stellar cluster, possibly of a flattened shape in its inner region \cite{kunneriath14,schoedel14}. The fall-back rate of the remnant debris onto SMBH is expected to be influenced by relativistic effects \cite{cheng14}.

The authors of ref.\ \cite{Clausen12} demonstrate that the photoionised debris of a tidally disrupted HB star can account for the emission lines observed in some optical spectra. In their case, the super-Eddington phase lasts about one to two hundred years; reproducing the line ratios requires an intermediate-mass black hole of $M_\bullet \lesssim 200 M_\odot$. Various characteristics of TDEs depend strongly on the stellar type \cite{macleod12}, although the gradual decay of the light curve adopts a generic profile that is determined by the viscous processes. The emission of an X-ray irradiated flow, known as the reflection spectrum, can be expected, including a superposition of the continuum emission and spectral lines, including the prominent fluorescent emission lines of iron that have the rest energy around 6--7 keV \cite{2006AN....327..961K,karas00,karas01,ross93}. In another context of X-ray emission lines, a method for O stars to determine the shock-heating rate by instabilities in their radiation-driven winds has been recently developed \cite{Cohen14}. For this paper, we just remind the reader that the relativistically smeared spectral line emission from a black-hole accretion disc in a few keV band may be intrinsically narrow and unresolved in energy; it is the observed profile that becomes broadened and skewed (by a combination of Doppler and light-bending effects) when integrated over the azimuthal extent of accretion rings, and generally red-shifted by strong gravity of the central black hole.

One of the most constraining X-ray spectral information on a tidal disruption event (TDE) has resulted from the campaign on a quiescent galaxy SDSS1201+30 \cite{saxton12}. This is most likely an object without a prior accretion disc. The X-ray spectra are very soft, and can hardly be explained with standard accretion disk models. The strong variability was seen in the light curve and related to clumpy accretion with a combination of flaring and absorption events.

X-ray photons from TDE can illuminate, ionise and perturb the gaseous material infalling from the fresh accretion flow. In ref.\ \cite{zhang14} we propose that the relativistic iron line from the TDE irradiated inflow can provide a unique way to determine parameters of the system, namely, the dimensionless black-hole spin $a$ ($-1\leq a \leq1$), the angle of the observer $i$ (inclination, $0\leq i\leq 90$ deg; $i=0$ corresponds to pole-on view along the rotation axis), and the expansion velocity $v_{\rm exp}$ of ionisation front, which propagates outwards at velocity close to the speed of light and modulates the ionisation parameter of the medium. Hence, it affects also the iron-line emissivity. 

According to the model, the line emission is triggered and modulated by the same, virtually instant TDE event. Unlike the case of continuously variable X-ray spectrum of active galactic nuclei \cite{esquej12,saxton12}, a quiescent source would be preferred, so that the transient relativistic line can be revealed. 

\section{Hypothesis of a relativistic spectral line from TDE-irradiated remnant accretion disc}
In ref.\ \cite{subr04} we modelled the structure of the nuclear star cluster that is expected to arise as a result of the mutual interaction of main components of the galactic nucleus, i.e., a supermassive black hole, stars, and a gaseous/dusty torus that helps to bring stars towards the tidal disruption. An interesting conclusion from that paper concerns the formation of a flattened (oblate) stellar subsystem in the inner region surrounding the central SMBH \cite{just12,Vilkoviskij02}, which is a reminiscence on much smaller scales of the structure originally reported on the kiloparsec scale of the bulge of the Milky Way \cite{launhardt02}. In the other words, TDE from such a stellar population can be expected to occur predominantly in the equatorial plane of the system. Therefore, also the irradiation flash should arise with a higher chance within the plane, i.e., at large inclination angle with respect to a distant observer. 

In our scenario for TDE, the flattened stellar system is associated with the effect of hydrodynamical dissipation of the stellar orbits by the dusty/gaseous torus. Further down, at very small radii near the horizon, also gravitational radiation can play a role. Moreover, a subset of stars on eccentric orbits is brought close to the central black hole where they plunge quickly below the tidal radius. Studying such a TDE would shed light on the accretion physics and the stellar dynamics in galactic nuclei, as well as the role of tidal disruptions for feeding and growth of SMBHs. 

We put forward a possibility that the source of intense energy deposition occurs near the inner edge of the accretion disc, which resembles a radially narrow ring that gradually spreads by viscous forces and eventually disappears once all material is accreted. Basic form of the relativistic line from a narrow ring were investigated in ref.\ \cite{sochora11}. Following TDE, the spectral line is modulated via changing ionisation state of a remnant accretion disc that is created from the debris. This remnant accretion ring can be identified with a structure that has been proposed to arise around the circularisation radius at transient accretion events with low (sub-Keplerian) angular momentum \cite{bu14,czerny13,hayasaki13}. 

\begin{figure}[tbh]
\centering
\includegraphics[width=0.6\textwidth]{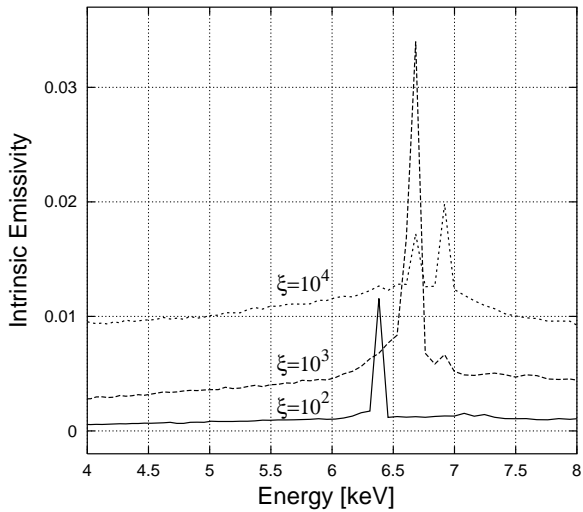}
\caption{The intrinsic emissivity profile (arbitrary units) around the 6--7\,keV spectral feature, as predicted by the model for hot parts of the ionisation front \cite{karas00}. The three curves correspond to different values of the ionisation parameter $\xi$. As the ionisation front expands through the infalling remnant disc, the changing ionisation state is revealed by the corresponding change of the iron line rest energy and intensity, until the emission feature disappears when the complete ionisation is reached.}
\label{fig1}
\end{figure}

A non-negligible radial infall velocity of the accreted material is likely, and so the ionisation front moves radially outward through the material of the remnant disc at $v=v_{\rm exp}(r)$ that is not exceeding the speed of light (although it should be very close to it). We can set $v_{\rm exp}\simeq {\rm const}$ as one of free model parameters. The corresponding mass fall-back rate proceeds as a characteristic power-law profile \cite{rees88},
$ \dot{M}(t)\propto K\;t^{-5/3}$
($K\equiv M_{\star}/t_{\rm fb}={\rm const}$), which can significantly exceed the Eddington accretion rate for a period of weeks to years for the black hole mass $M_{\bullet} \lesssim 10^7 ~\rm M_\odot$
\cite{strubbe09}. 

\begin{figure}[tbh]
\centering
\includegraphics[width=0.99\textwidth]{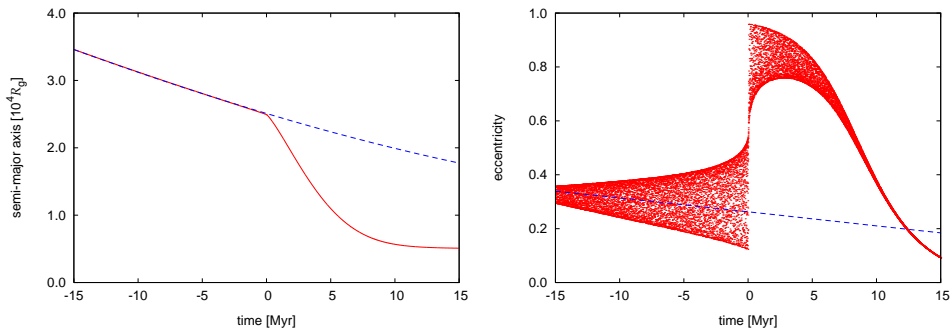}
\caption{Temporal evolution of semi-major axis and eccentricity of the orbit of a star from a nuclear star cluster interacting with a disc \cite{subr05b}. Dashed line corresponds to the case when the effect of disc gravity is neglected. Solid line (in the left panel) and dots (in the right panel) represent a simulation where both the hydrodynamical and gravitational interaction were considered. According to this scheme, TDE event is triggered at the moment of close encounter (zero time on the horizontal axis) between the star and SMBH ($M_\bullet\simeq4\times 10^6M_\odot$).}
\label{fig2}
\end{figure}

The hypothesis about an enhanced capture rate and tidal disruptions of stars in the equatorial plane is supported also by simulations of the structure flattening in the inner regions of the nuclear cluster. In ref.\ \cite{subr05} a scenario was discussed based on a combination of simultaneous gravitational and hydrodynamical effects of the gaseous environment on orbiting stars, which are assumed to lead to an oblate, disc-like configuration. While the direct star-disc hydrodynamical interaction causes a continuous dissipation of the stellar orbital energy (and it is anyway very small in the Milky Way's central regions), gravity of the stellar ring, a self-gravitating accretion torus, or a flattened nuclear star-cluster all can lead to periodical variations of the orbital elements. An example of this evolution is plotted in Figure \ref{fig2}, where we show the long-term orbital changes due to both the hydrodynamical and gravitational influence of the disc. As an example, parameters of that simulation were set  to be consistent with the values reported for S2 star in the Galactic center. Alternatively, tidal disruption of stars by supermassive central black holes from dense star clusters has been modelled by high-accuracy direct N-body simulations \cite{zhong14}.

We note that in the X-ray and optical/UV bands, almost two dozens candidate TDEs have been already reported \citep{Gezari:2012fk,Greiner:2000vx,Komossa:1999uy}. These events are characterized by the thermal emission with temperature of $\sim 10^4$--$10^5$~K, and the peak bolometric luminosity about $10^{43}$--$10^{45}~{\rm erg~s^{-1}}$. For the events with good coverage during the decay, the flux decline was found consistent with a power-law (index $-5/3$), as predicted for canonical TDEs. Compared with the quiescence state, the flux can increase by a factor of $\gtrsim150$ during outbursts.

Detections of the fluorescent iron line from TDE are only tentative so far \cite{evans89}, however, finding this spectral signature should provide very valuable information. The expected properties of the reflection line depend sensitively on the ionisation state of the irradiated material \cite{ross93}. Since in TDEs the continuum flux can vary by a large amplitude, we need to take also the variation of the ionisation parameter $\xi(t)$ into consideration (see Figure~\ref{fig1}). For $M_\bullet\approx10^6~\rm M_{\odot}$ SMBH, the peak mass fall-back rate can reach $\sim 1.5 ~\rm M_{\odot}~yr^{-1}\!$, corresponding to luminosity of $8.5\times10^{45}~\rm erg~s^{-1}$ (assuming the radiation efficiency of $\eta\simeq0.1$). Therefore, the rise of the illumination is significant by orders of magnitude compared to the quiescent state of a dormant nucleus.

\section{Discussion and Conclusions}
Gravitational effects act on the spectral features from the remnant TDE accretion disc by smearing the spectral features and moving the observed energy centroid across energy bins. In this way gravity exerts the influence on the ultimate form of the spectrum \citep{2006AN....327..961K}. The reprocessed radiation reaches the observer from different regions of the system. Furthermore, as strong-gravity plays a crucial role, photons may even follow multiple separate paths, joining each other at the observer at different moments. Individual rays experience unequal time lags for purely geometrical reasons and for relativistic time-dilation.

Time delay from a TDE flare to the moment of arrival of the observed iron line signal consists of two components: the expansion time $t_{\rm exp}=r/v_{\rm exp}$, and the time interval from the disc to the observer, $t_{\rm delay}$. At large radius where the spacetime is flat to a good precision, $t_{\rm delay}\simeq -r\sin\phi+\mbox{const}$, where $\phi$ is the azimuthal angle on the disc plane \cite{karas01}. As $v_{\rm exp}$ is of order unity (i.e., comparable to the speed of light),  the two quantities become comparable, and the delay interval is dominated by the longer one. At large inclination angles, the photons from the disc located in front of the black hole reach the observer first, then those from disc near the inner edge, and at last those from disc behind black hole.

In the cases of high expansion velocity combined with high inclination angle, a ``nose'' occurs ahead of the rings main signal \cite{zhang14}. The length of the nose increases with the inclination angle and with the expansion velocity. The line emission ceases once the accretion flow becomes highly ionised to a larger distance.  On a phenomenological level of our model, the intrinsic emissivity of the remnant accretion ring and the expansion velocity of the ionisation front are two degrees of freedom that allow us to capture the lamp-post scheme as well as the TDE-induced illumination in a common scenario. We note that the remnant accretion rings resulting from a TDE event are likely lacking axial symmetry. Therefore, as a next step to explore more realistic situations, spectral line profiles from elliptically shaped and tilted structures need to be taken into account \cite{chang02,Eracleous95,Fragile05}.

We conclude by stating that the proposed idea poses an observational challenge (because the iron line flux is expected to be only a weak and variable components of the X-ray signal from tidal-disruption events), nevertheless, it suggests a promising opportunity to verify the parameters of central black holes by an independent method. Moreover, it offers and interesting complement to the standard scenario for the origin of relativistically broadened spectral lines. 

In the context of Galactic centre, let us remark that the 6--7 keV emission of iron has been extensively studied \cite{ponti10,wang06} and the light-echo effect reported. However, the detection concerns a wider region (molecular clouds within the Central Molecular Zone surrounding the SMBH) than the immediate vicinity of the black-hole horizon, which we imagine here.

While in the lamp-post model the irradiating source is (usually) considered to be (almost) axially symmetric and located around the rotation axis, in the present scheme the primary excitation is more likely to occur from the equatorial region near the innermost stable circular orbit, commonly denoted as ISCO, which is about the minimum radius to which the accretion discs can extend. A convincing case of such a transient relativistic spectral feature from TDE is still to be found in X-rays.

\ack
The authors highly appreciate the hospitality of Conference organisers in Prague. VK and DK thank the Czech Ministry of Education, Youth and Sports project LH14049, titled ``Spectral and Timing Properties of Cosmic Black Holes'', that has been aimed to support the international collaboration in astrophysics. VK and MD acknowledge also the European Union 7th Framework Programme No. 312789  ``StrongGravity''. The research of WY and WZ is supported in part by the National Natural Science Foundation of China under grants No.\ 11073043 and No.\ 11333005.



\end{document}